\documentclass[12pt]{iopart}

\usepackage{color}
\usepackage{url}
\usepackage[dvips]{graphicx}

\newcommand{\fourier}[1]{\mathcal{F} \biggl[#1 \biggl]}
\newcommand{\ifourier}[1]{\mathcal{F}^{-1} \biggl[#1 \biggl]}

\newcommand{\pd}[2]{ \frac{\partial #1}{\partial #2} }
\newcommand{\absbar}[1]{\left |#1 \right |} 

\begin{document}

\title{Aliasing-reduced Fresnel diffraction with scale and shift operations}

\author{Tomoyoshi Shimobaba,$^{1}$ Takashi Kakue,$^1$ Naohisa Okada,$^1$ Minoru Oikawa,$^1$ Yumi Yamaguchi,$^1$ and Tomoyoshi Ito$^{1}$}

\address{
$^1$Graduate School of Engineering, Chiba University, 1-33 Yayoi-cho, Inage-ku, Chiba 263-8522, Japan
}
\ead{shimobaba@faculty.chiba-u.jp}
\begin{abstract}

Numerical simulation of Fresnel diffraction with fast Fourier transform (FFT) is widely used in optics, especially computer holography. 
Fresnel diffraction with FFT cannot set different sampling rates between source and destination planes, while shifted-Fresnel diffraction can set different rates.
However, an aliasing error may be incurred in shifted-Fresnel diffraction in a short propagation distance, and the aliasing conditions have not been investigated. 
In this paper, we investigate the aliasing conditions of shifted-Fresnel diffraction and improve its properties based on the conditions.

\end{abstract}


\maketitle

\section{Introduction}
Numerical implementation of Fresnel diffraction \cite{goodman} with fast Fourier transform (FFT) is widely used in optics, especially computer-generated holograms (CGH) and digital holography \cite{poon,dh}. 
CGH is a technique for generating a hologram pattern on a computer by simulating diffracted light from objects, and its applications include three-dimensional display and the design of optical elements.
Digital holography reconstructs the light of an object including the amplitude and phase on a computer from a hologram captured by imaging devices such CCD cameras.

In these fields, Fresnel diffraction and the angular spectrum method are used as key techniques.
These diffraction calculation can be accelerated by FFT-implemented convolution; however it imposes the limitation that the sampling rate on a source plane is the same as that on a destination plane due to the property of FFT.
If we can change the sampling rate, we will be able to extend the applications of diffraction calculations. 

In the angular spectrum method, some studies address different sampling rates on source and destination planes.
Reference \cite{sasm1} proposed the scaled angular spectrum method using scaled-FFT, which is one version of FFT introducing a scaling parameter.
This angular spectrum method requires four FFTs.
Another one was proposed by us \cite{sasm2}.
Our scaled angular spectrum method was implemented by non-uniform FFT.
This angular spectrum method requires two FFTs and one interpolation.

Fresnel diffraction to address the scale operation has already been realized by some studies. 
Double-step Fresnel diffraction executes the scale operation by double Fresnel diffraction calculations \cite{yamaguchi}. 
Shifted-Fresnel diffraction addresses the shift and scale operations by scaled-FFT \cite{shift1}. 
Reference \cite{shift2} proposed the same method as the shifted-Fresnel diffraction by Bluestein transform.
Scaled-FFT and Bluestein transform are essentially the same.

In holography, the scale operation of shifted-Fresnel diffraction \cite{shift1} is used in several applications: for example, wavefront-recording method \cite{wrp1,wrp2,wrp3} for accelerating CGH generation, digital holographic microscopy observable at any magnification \cite{dhm}, and zoomable holographic projection without a zoom lens \cite{proj}.
Shifted-Fresnel diffraction yields new applications; however, the calculations may incur a serious aliasing error in a short propagation distance and the aliasing error has not been investigated. 

In this paper, we investigate the aliasing conditions of shifted-Fresnel diffraction and improve the problems incurred based on the conditions.
The improved diffraction is referred as to aliasing-reduced shifted and scaled (ARSS)-Fresnel diffraction.
In Section 2, we explain ARSS-Fresnel diffraction.
In Section 3, we show a comparison of diffracted results between ARSS-Fresnel diffraction and shifted-Fresnel diffraction.
Section 4 concludes this work.

\section{Aliasing-reduced Fresnel diffraction with the scale and shift operations}

Fresnel diffraction is expressed by, 
\begin{eqnarray}
u_2(x_2) =  \frac{\exp(i k z)}{i \lambda z}  \int u_1(x_1) 
\exp( \frac{i \pi}{\lambda z} (x_2-x_1)^2 ) dx_1 
\label{eqn:fre}
\end{eqnarray}
where $k$ is the wave number, $\lambda$ is the wavelength of light, $u_1(x_1)$ is the source plane, $u_2(x_2)$ is the destination plane,  and $z$ is the propagation distance between the source and destination planes.

We derive ARSS-Fresnel diffraction by the following relation:
\begin{equation}
(x_2-s x_1+o)^2=s(x_2-x_1)^2+(s^2-s)x_1^2+(1-s)x_2^2+2 o x_2 -2 s o x_1 + o^2
\label{eqn:relation}
\end{equation}
where $s$ is the scaling parameter, $x_1=p m_1$ and $x_2=p m_2$.
$m_1$ and $m_2$ are integers with ranges of $m_1, m_2 \in [-N/2, N/2-1]$ where $N$ is the number of sampling points.
$p$ is the sampling rate on the destination plane and $o$ is offset from the origin.
The sampling rate on the source plane is determined by $s p$.

We can obtain the following equation by substituting the relation into Eq.(\ref{eqn:fre}): 
\begin{eqnarray}
u_2(x_2) &=&  C_z \int u_1(x_1) \exp(\frac{i \pi}{\lambda z}((s^2-s)x_1^2-2sox_1)) \exp(\frac{i \pi s(x_2-x_1)^2}{\lambda z}) dx_1 \nonumber \\ 
\label{eqn:prev_arss}
\end{eqnarray}

We can obtain ARSS-Fresnel diffraction by using the convolution theorem and introducing the band-limiting function to reduce aliasing:  
\begin{eqnarray}
u_2(x_2) &=&  C_z \ifourier{ 
	\fourier{u_1(x_1) \exp(i \phi_u)} 
	\fourier{\exp(i \phi_h) {\rm Rect}(\frac{x_h}{2 x_{max}}) }  }
\label{eqn:arss}
\end{eqnarray}
where ${\rm Rect}(\cdot)$ is the rectangular function, $x_{h}=p m_{h}$ ( $m_h \in [-N/2, N/2-1]$), and $\exp(i \phi_u)$, $\exp(i \phi_h)$ and $C_z$ are defined by,
\begin{eqnarray}
\exp(i \phi_u) = \exp(i \pi \frac{(s^2-s)x_1^2-2sox_1}{\lambda z}) \\
\exp(i \phi_h) = \exp(i \pi \frac{sx_h^2}{\lambda z}) \\
C_z=\frac{\exp(i \phi_c)}{ i \lambda z }=\frac{\exp(i kz + \frac{i \pi}{\lambda z}((1-s)x_2^2+2ox_2+o^2))}{i \lambda z} 
\label{eqn:cz}
\end{eqnarray}

Band-limitation function will be acceptable other than the rectangular function, for example, the circular function, window functions (e.g. Hanning and Hamming window functions) and so forth.  
It is straightforward way to extend to two-dimensional ARSS-Fresnel diffraction because Fresnel diffraction allows separation of variables.

\begin{figure}[htb]
\centerline{
\includegraphics[width=12cm]{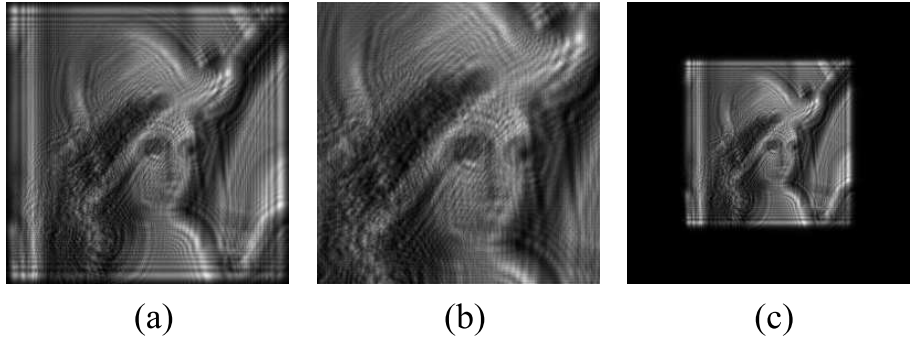}}
\caption{Light intensity distributions calculated by shifted-Fresnel diffraction \cite{shift1,shift2}. (a) $s=1$ (b) $s=6/4$ (c) $s=6/10$.}
\label{fig:sfre-ex}
\end{figure}

Figure \ref{fig:sfre-ex} shows the light intensity distributions calculated by shifted-Fresnel diffraction \cite{shift1,shift2}. 
The calculation parameters are $N=1,024$, $\lambda=633$ nm and $z=0.1$ m.
The sampling rates on the source and destination planes indicate $p_1$ and $p_2$, respectively. 
The scaling parameter $s (=p_1/p_2)$ of Figs \ref{fig:sfre-ex} (a), (b) and (c)  are $s=1$($p_1$ and $p_2$ are 6 $\mu$ m), $s=6/4$ and $s=6/10$, respectively.

\begin{figure}[htb]
\centerline{
\includegraphics[width=14cm]{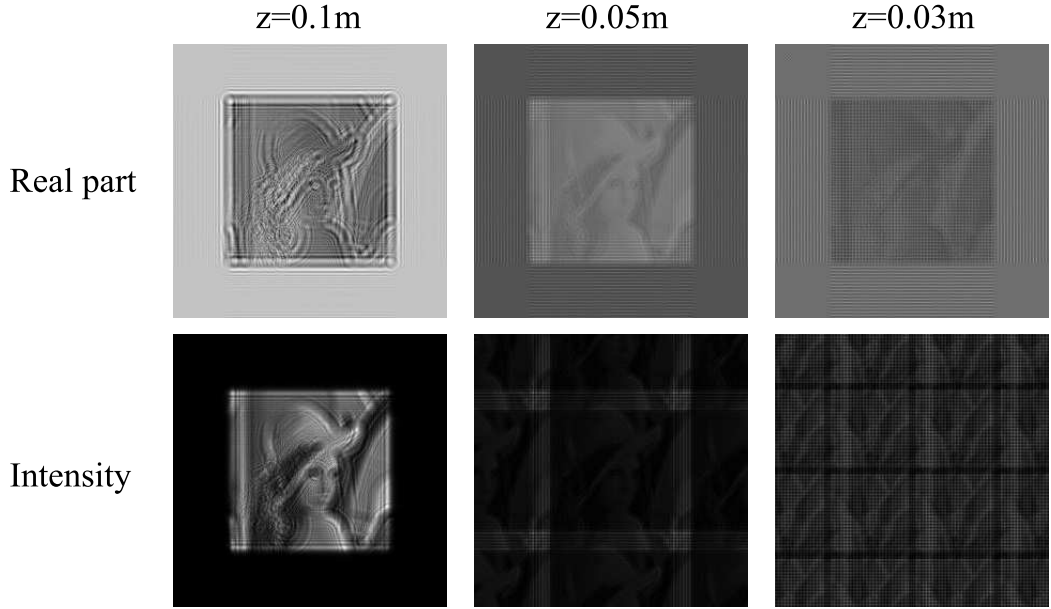}}
\caption{Real part and intensity of shifted-Fresnel diffraction at the propagation distance of 0.1 m, 0.05 m and 0.03 m.}
\label{fig:sfre-aliasing}
\end{figure}

As shown in the figure, shifted-Fresnel diffraction can magnify diffracted results by the scaling parameter; however, no countermeasure for aliasing has been devised.
For instance, when changing $z$ from 0.1 m to 0.05 m and 0.03 m, the results are shown in Fig.\ref{fig:sfre-aliasing}.
The upper images show the real parts and the bottom images show the light intensity of shifted-Fresnel diffraction. 
In the case of $z=0.1$m, the results have good quality without aliasing.
However, in the cases of $z=0.05$m and $z=0.03$m, the results are extremely contaminated by aliasing.

\subsection{Aliasing conditions}

\begin{figure}[htb]
\centerline{
\includegraphics[width=14cm]{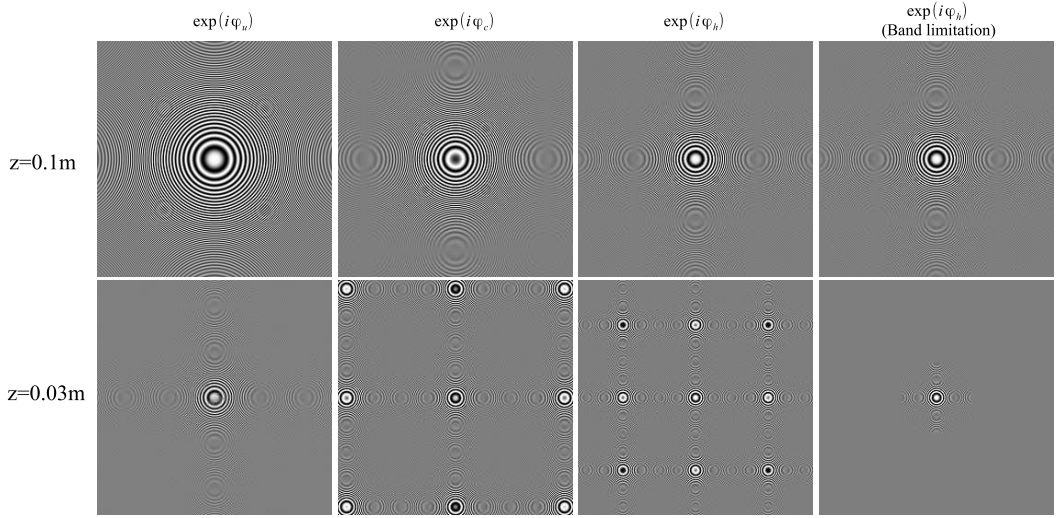}}
\caption{Real part of the chirp functions at the propagation distance of $z=0.1$ m and $z=0.03$ m. }
\label{fig:zone}
\end{figure}

Although ARSS-Fresnel diffraction is derived in a different way to shifted-Fresnel diffraction, ARSS-Fresnel diffraction of Eq.(\ref{eqn:arss}) is essentially the same as shifted-Fresnel diffraction, except for the introduction of the band-limitation function. 
In this subsection, we clarify the aliasing conditions of ARSS-Fresnel diffraction. 
Equation (\ref{eqn:arss}) involves three chirp functions, namely $\exp(i  \phi_c)$, $\exp(i \phi_u)$ and $\exp(i \phi_h)$.
The cause of aliasing is the chirp functions.
Figure \ref{fig:zone} shows the real part of the chirp functions, under the calculation conditions of $N=1,024$, $o=0$, $\lambda=633$ nm, $p=10 \mu$m and $s=6/10$. 
In the case of $z=0.1$ m, the aliasing is not incurred in the chirp functions; however, in the cases of $z=0.03$ m, $\exp(i \phi_c)$ and $\exp(i \phi_h)$ are contaminated by aliasing.

Therefore, we need to band-limit these functions.
The aliasing-free areas of the chirp functions can be calculated by \cite{goodman},
\begin{equation}
\frac{1}{p} \geq 2\absbar{f_c} = 2 \absbar{\frac{1}{2 \pi} \pd{\phi_c}{x_2} }=\absbar{ \frac{2(1-s)x_2+2so}{\lambda z} }
\label{eqn:freq_cz}
\end{equation}
\begin{equation}
\frac{1}{p} \geq 2|f_u| = 2 \absbar{ \frac{1}{2 \pi} \pd{\phi_u}{x_1} } =
\absbar{ \frac{2(s^2-s)x_1-2so}{\lambda z} }
\label{eqn:freq_u1}
\end{equation}
\begin{equation}
\frac{1}{p} \geq 2|f_h| = 2 \absbar{ \frac{1}{2 \pi} \pd{\phi_h}{x_h} } = 
\absbar{ \frac{2 s x_h}{\lambda z} }
\label{eqn:freq_h}
\end{equation}

From Eqs.(\ref{eqn:freq_cz})-(\ref{eqn:freq_h}), the aliasing-free areas in pixel units are as follows:
\begin{equation}
\absbar{ m_2 } \leq  \absbar{ \frac{\lambda \absbar{z}  -\absbar{2so}  }{2(1-s) p^2}}
\label{eqn:r_cz}
\end{equation}
\begin{equation}
\absbar{ m_1 } \leq  \absbar{ \frac{\lambda \absbar{z}  -\absbar{2so}  }{2(s^2-s) p^2}}
\label{eqn:r_u1}
\end{equation}
\begin{equation}
\absbar{ m_h } \leq  \absbar{ \frac{\lambda z  }{2s p^2}}
\label{eqn:r_h}
\end{equation}

According to Eq.(\ref{eqn:r_h}), the band-limited chirp function of $\exp(i  \phi_h)$ at $z=0.03$m  is shown in Fig.\ref{fig:zone}.

\begin{figure}[htb]
\centerline{
\includegraphics[width=17cm]{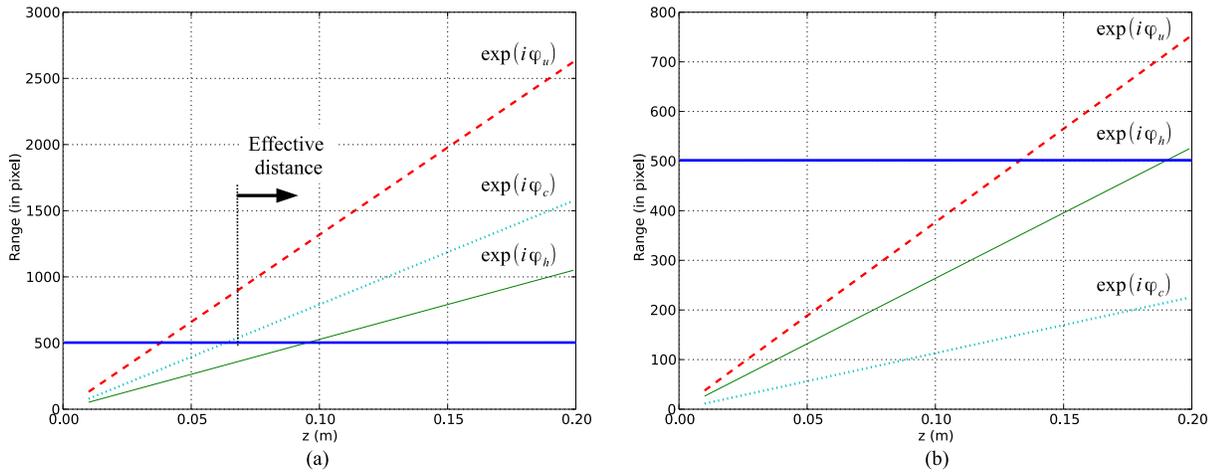}}
\caption{Aliasing-free area in pixel units of the chirp functions under the calculation condition of $\lambda=633$ nm. (a) $s=6/10$ ($p=10 \mu$m) (b) $s=6/20$ ($p=20 \mu$m).}
\label{fig:plot1}
\end{figure}

Changing $z$ between 0.01 m and 0.25 m,  Fig.\ref{fig:plot1} plots the aliasing-free areas in pixel units of the chirp functions under the calculation conditions of $N=1,024$, $o=0$, $\lambda=633$ nm, $p=10 \mu$m and $s=6/10$.
In the figure, the green, red, and light blue plots indicate
 the aliasing-free areas of $\exp(i \phi_h)$, $\exp(i \phi_u)$ and $\exp(i  \phi_c)$, respectively.
The blue line indicates the half area ($N/2$) of the source and destination planes.
The aliasing is incurred when each plot falls below the blue line. 

In Fig.\ref{fig:plot1}(a), the effective distance without aliasing is over about 0.07 m because the chirp functions $\exp(i \phi_c)$ and $\exp(i  \phi_u)$ do not incur aliasing over about 0.07 m, and we need to band-limit $\exp(i \phi_h)$  in about 300 pixels ($x_{max} \approx 300 p$ in Eq.(\ref{eqn:arss})).

In Fig.\ref{fig:plot1}(b), the chirp functions $\exp(i \phi_u)$ and $\exp(i  \phi_h)$ for $p=20 \mu$m and $s=6/20$ do not incur aliasing over about 0.18 m, while $\exp(i \phi_c)$ still incurs aliasing at about 0.18 m.
If we band-limit $\exp(i \phi_c)$ in the aliasing-free area of about 200 pixels, no aliasing is incurred; however, the complex amplitude resulting in Eq.(\ref{eqn:arss}) outside the aliasing-free area was eliminated because zero area of $\exp(i \phi_c)$ outside the aliasing-free area is multiplied with the result of inverse FFT of Eq.(\ref{eqn:arss}).

Therefore, for complex amplitude, we need to satisfy the following relation to avoid aliasing:
\begin{equation}
\absbar{m_1}, \absbar{m_2} \geq N/2
\label{eqn:aliasing_complex}
\end{equation}

For light intensity, we need to satisfy the following relation to avoid aliasing:
\begin{equation}
\absbar{m_1} \geq N/2
\label{eqn:aliasing_intensity}
\end{equation}
because the $\exp(i  \pi \phi_c)$ can be ignored in the light intensity.
The aliasing condition of the light intensity $\absbar{u_2(x_2)}^2$ is mitigated compared to that of complex amplitude.

\begin{figure}[htb]
\centerline{
\includegraphics[width=17cm]{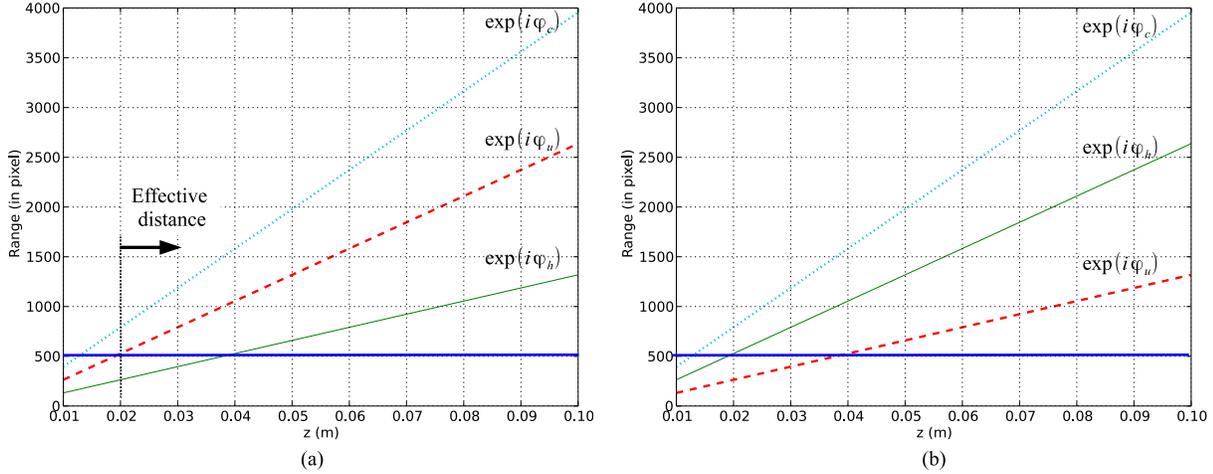}}
\caption{Aliasing-free area in pixel units of the chirp functions under the calculation condition of $\lambda=633$ nm. (a) $s=6/4$ ($p=4 \mu$m) (b) $s=6/2$ ($p=2 \mu$m).}
\label{fig:plot2}
\end{figure}

In Fig.\ref{fig:plot2}(a), the effective distance without aliasing in the scaling parameter of $s=6/4$ ($p=4 \mu$m) is over about 0.02 m because the chirp functions $\exp(i \phi_c)$ and $\exp(i \phi_u)$ do not incur aliasing over about 0.02 m, and we need to band-limit $\exp(i \phi_h)$  in the aliasing-free area of about 250 pixels  ($x_{max} \approx 250 p$ in Eq.(\ref{eqn:arss})).
In Fig.\ref{fig:plot2}(b), the chirp functions $\exp(i  \pi \phi_c)$ and $\exp(i  \pi \phi_h)$ in the scaling parameter of $s=6/2$  ($p=2 \mu$m) do not incur aliasing over 0.02 m, while $\exp(i \phi_u)$ incurs aliasing at the distance.
If we band-limit $\exp(i \phi_u)$ in the aliasing-free area of about 250 pixels, the aliasing is not incurred; however, the complex amplitude and intensity of Eq.(\ref{eqn:arss}) was eliminated outside the 250 pixels.

\section{Results}

\begin{figure}[htb]
\centerline{
\includegraphics[width=13cm]{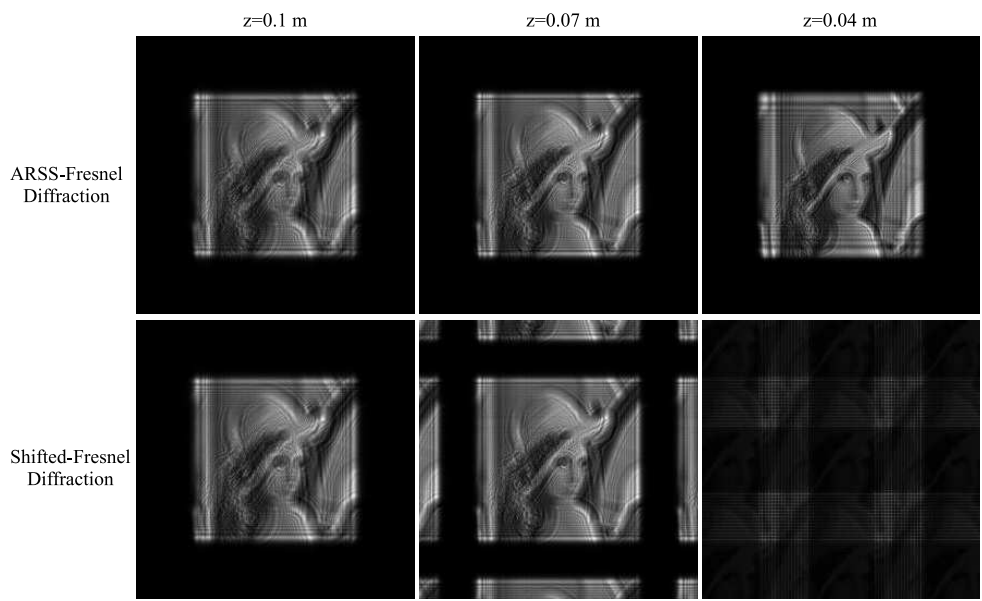}}
\caption{Intensity distributions calculated by ARSS-Fresnel diffraction and shifted-Fresnel diffraction. The scaling parameter is $s=6/10$.}
\label{fig:result1}
\end{figure}

Figure \ref{fig:result1} shows the intensity distributions calculated by ARSS-Fresnel diffraction and shifted-Fresnel diffraction \cite{shift1,shift2}, under the calculation conditions of $N=1,024$, $\lambda=633$ nm, $p=10 \mu$m, and $s=6/10$.
We change the propagation distance from $z=0.1$ m to 0.07 m and 0.04 m.
As we can see, the results of ARSS-Fresnel diffraction do not incur aliasing, while those of shifted-Fresnel diffraction incur aliasing at $z=0.07$ m and $0.04$ m.

\begin{figure}[htb]
\centerline{
\includegraphics[width=13cm]{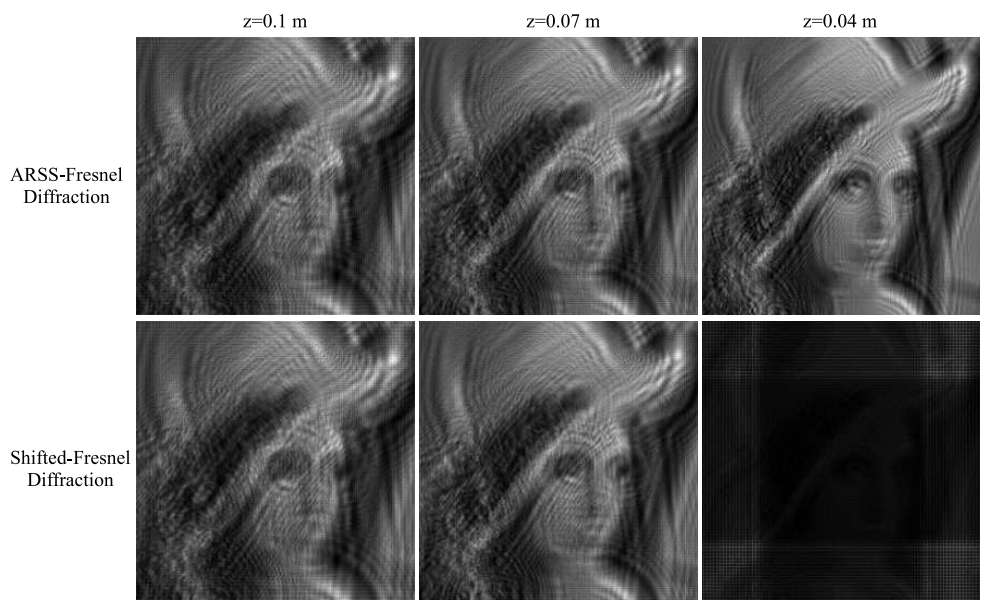}}
\caption{Intensity distributions calculated by ARSS-Fresnel diffraction and shifted-Fresnel diffraction. The scaling parameter is $s=6/4$.}
\label{fig:result2}
\end{figure}

Figure \ref{fig:result2} shows the intensity distributions calculated by ARSS-Fresnel diffraction and shifted-Fresnel diffraction, under the calculation conditions of $\lambda=633$ nm, $p=4 \mu$m, and $s=6/4$.
As we can see, the results of ARSS-Fresnel diffraction do not incur aliasing, while those of shifted-Fresnel diffraction incur aliasing at $0.04$ m.
From Figs.\ref{fig:result1} and \ref{fig:result2}, ARSS-Fresnel diffraction can be applied to a wider propagation distance than shifted-Fresnel diffraction. 

\section{Conclusion}
We clarified the aliasing conditions of shifted-Fresnel diffraction and improved the diffraction, which was named ARSS-Fresnel diffraction.
ARSS-Fresnel diffraction is useful for CGH calculation and digital holography because of the scale and shift property.

\section*{Acknowledgement}
This work is supported by Japan Society for the Promotion of Science (JSPS) KAKENHI (Grant-in-Aid for Scientific Research (C) 25330125) 2013, and KAKENHI (Grant-in-Aid for Scientific Research (A) 25240015) 2013.

\end{document}